# Angstrom-Resolution Magnetic Resonance Imaging of Single Molecules via Wavefunction Fingerprints of Nuclear Spins

Wen-Long Ma and Ren-Bao Liu[*]

*Department of Physics and Centre for Quantum Coherence, The Chinese University of Hong Kong, Shatin, New Territories, Hong Kong, China*

**Abstract**

Single-molecule sensitivity of nuclear magnetic resonance (NMR) and angstrom resolution of magnetic resonance imaging (MRI) are the highest challenges in magnetic microscopy. Recent development in dynamical decoupling (DD) enhanced diamond quantum sensing has enabled single-nucleus NMR and nanoscale NMR. Similar to conventional NMR and MRI, current DD-based quantum sensing utilizes the "frequency fingerprints" of target nuclear spins. The frequency fingerprints, by their nature, cannot resolve different nuclear spins that have the same noise frequency or differentiate different types of correlations in nuclear spin clusters, which limit the resolution of single-molecule MRI. Here we show that this limitation can be overcome by using "wavefunction fingerprints" of target nuclear spins, which is much more sensitive than the "frequency fingerprints" to the weak hyperfine interaction between the targets and a sensor under resonant DD control. We demonstrate a scheme of angstrom-resolution MRI that is capable of counting and individually localizing single nuclear spins of the same frequency and characterizing the correlations in nuclear spin clusters. A nitrogen-vacancy centre spin sensor near a diamond surface, provided that the coherence time is improved by surface-engineering in the near future, may be employed to determine, with angstrom-resolution, the positions and conformation of single molecules that are isotope-labelled. The scheme in this work offers an approach to breaking the resolution limit set by the "frequency gradients" in conventional MRI and to reaching the angstrom-scale resolution.

---

[*] Corresponding author. Email: rbliu@phy.cuhk.edu.hk

# I. Introduction

Detection of single nuclear spins has broad applications, such as molecular structure analysis in chemistry and biology [1] and spin-based quantum computing [2]. However, the weak signals from single nuclear spins cannot be detected by conventional nuclear magnetic resonance (NMR) [3]. Recently the nitrogen-vacancy (NV) center in diamond, for its long coherence time at room temperature [4-6], has been used to detect AC magnetic fields [7-8], magnetic noises [9], single nuclear spins [10-14], and nanoscale nuclear spin ensembles [15-18] via dynamical decoupling (DD) enhanced quantum sensing. The principle of DD-enhanced quantum sensing is to identify the "frequency fingerprints" of target nuclear spins [19, 20]. When the frequency of a DD sequence on the quantum sensor (e.g., the electron spin of an NV center) matches the transition frequency of a target nuclear spin or a spin cluster, i.e., when the resonant DD condition is realized, the noise from the target nuclear spins is resonantly enhanced and hence the sensor coherence presents a sharp dip. The "frequency fingerprints" are also the basis of conventional NMR and MRI. In particular, the resolution of MRI relies critically on the frequency gradients (which is proportional to the magnetic field gradients) of nuclear spins.

The application of DD-enhanced quantum sensing to magnetic resonance imaging (MRI) of single molecules, however, is restricted by two critical issues. First, the DD-based quantum sensing cannot distinguish target nuclear spins that have the same "frequency fingerprints". For example, the target spins of the same species, when weakly coupled to the sensor, would precess with the same Larmor frequency under a uniform external magnetic field and cannot be individually resolved. A remarkable solution is to induce nanoscale magnetic field gradients by using nanoscale magnetic tips [21, 22] or using the anisotropic hyperfine interaction with an electron spin sensor [18], which produces a nanoscale frequency gradient for target nuclear spins. Nanoscale MRI has been achieved this way [1, 15-18, 23, 24]. However, to push the MRI resolution to angstrom-scale, the field gradient due to a sensor spin [18] or a magnetic nano-tip [21] is still too small (typically 0.8~0.1 G nm$^{-1}$ for the distance 3~5 nm) [25]. Previous works have studied the sensor coherence dips caused by a nuclear spin ensemble [15-17] and obtained spatial distributions of nuclear spins by numerical fitting [18]. There is still no scheme to individually identify nuclear spins of the same frequency. Furthermore, the DD-based quantum sensing cannot distinguish the correlated transitions and independent multiple transitions in a nuclear spin cluster if the two types of transitions have the same "frequency fingerprints". Therefore new schemes are still needed for quantum sensing to reach the same level of sophistication as the conventional multi-dimensional NMR [3].

Here, we propose a scheme of angstrom-resolution MRI of single molecules to overcome the limitations of the frequency fingerprint sensing schemes. Our new scheme is inspired by the quantum decoherence of sensor spins caused by weakly coupled target spins: Even if the target spins have the same "frequency fingerprints", they may have characteristically different quantum evolutions conditioned on the sensor state, presenting different "wavefunction fingerprints" onto the sensor spin decoherene. Before showing the results in details, we explain the main idea of wavefunction fingerprint MRI as follows. When the sensor spin is under periodic DD control, the periodically modulated hyperfine interaction amounts to oscillating magnetic fields on the target nuclear spins. The quantum evolution of the nuclear spins driven by the effective AC magnetic fields is conditioned on the sensor state $|\pm\rangle$, which records the which-way information of the sensor spin and hence breaks the phase correlation between the sensor spin states $|+\rangle$ and $|-\rangle$. In particular, under the resonant DD condition, the target spins perform resonant Rabi oscillations driven by the modulated hyperfine interaction. The frequencies of the Rabi oscillations are sensitive to the hyperfine interaction; in contrast the nuclear spin transition frequencies are insensitive to weak hyperfine interactions. The Rabi oscillations, i.e., the wavefunction evolutions of the target nuclear spins, induce sensor coherence dip oscillations with features characteristic of the weak hyperfine interactions. These sensor coherence oscillation features define the "wavefunction fingerprints" of the target spins. By their "wavefunction fingerprints", single nuclear spins and transitions in single nuclear spin clusters, even if they have the same frequencies, can be identified by the coherence dip oscillations of a quantum sensor (such as an NV center spin in diamond) as a function of the DD pulse number. The features of such oscillations determine the number of nuclear spins, the individual hyperfine coupling strengths, and the types of correlations, in a discrete manner: (1) The number of coherence zeros within a certain evolution time equals the number of nuclear spins of the same species coupled to the sensor with strengths above a certain threshold; (2) The positions of the zeros determine the individual strengths of the couplings between the nuclear spins and the sensor; (3) The maximum depth of the sensor coherence dip, taking discrete values (such as -1, -1/3, 0, 1/5, etc), determines the types of correlations in a nuclear spin cluster. Therefore, we can determine the number and positions of nuclear spins of the same species weakly coupled to a quantum sensor. For example, an NV center near a diamond surface, provided that the coherence time can be improved by surface engineering in the near future, can be employed to identify with angstrom-resolution the positions and conformations of single molecules (such as proteins) that are isotope labeled.

## II. "Frequency fingerprints" and "wavefunction fingerprints" of target spins

In this section, we provide the physical picture for identifying the "frequency fingerprints" and "wavefunction fingerprints" of target spins onto the coherence of a quantum sensor under DD control. The DD control of the sensor consists of a sequence of $\pi$-flips at times $\{t_1, t_2 \cdots t_N\}$ for the sensor evolution from 0 to $t$. The DD control suppresses the background environmental noise and selectively enhances the noise from the target spins [8, 10, 20]. In this paper we consider the *N*-pulse Carr-Purcell-Meiboom-Gill (CPMG-*N*) control [26, 27] with $t_p = (2p-1)\tau$ (where $2\tau$ is the interval between pulses and $p = 1, 2, \cdots, N$), as shown in Fig. 1(a).

Let us first consider a sensor spin (*S*=1/2) weakly coupled to a remote target spin (*I*=1/2). With the sensor spin under DD control, the total spin Hamiltonian (in the toggling reference frame rested on the free sensor under an ideal DD control) is

$$H(t) = f(t) S_z \otimes (\mathbf{A} \cdot \mathbf{I}) + \omega_0 I_z, \qquad (1)$$

where $f(t)$ is the DD modulation function jumping between +1 to -1 every time the sensor spin is flipped by a DD pulse [Fig. 1(b)], $\mathbf{A} = (A_x, A_y, A_z)$ is the hyperfine field of the target spin. We assume the weak coupling condition, i.e., $A \ll \omega_0$. For periodic DD control, $f(t) = f(t+T)$ with $T$ denoting the period, so the modulation function can be expanded into a Fourier series as $f(t) = \sum_{j=0}^{\infty} f_j \cos(j\omega_T t)$ with $\omega_T = 2\pi/T$ and $f_j = \frac{2}{T}\int_0^T f(t)\cos(j\omega_T t)dt$. For CPMG control, $T = 4\tau$ with $2\tau$ denoting the pulse interval, and the non-zero Fourier coefficients are $f_{(2q-1)} = 4(-1)^{q+1}/[(2q-1)\pi]$ with $q = 1, 2, \cdots$ [Fig. 1(c)], so the Hamiltonian in Eq. (1) conditioned on the sensor states $|\pm\rangle$ can be rewritten as

$$H^{(\pm)}(t) = \pm \sum_{q=1}^{\infty} \frac{2(-1)^{q+1}\cos[(2q-1)\omega_T t]}{(2q-1)\pi}(\mathbf{A}\cdot\mathbf{I}) + \omega_0 I_z, \quad (2)$$

which shows that the periodic flips of the sensor spin effectively impose on the target spin a series of oscillating magnetic fields with frequencies $(2q-1)\omega_T$ and corresponding field strengths $\sim A/(2q-1)$. The oscillating magnetic fields contain longitudinal components $\sim A_z I_z$ and transverse components $\sim (A_x I_x + A_y I_y)$. The transverse components drive the sensor spin into Rabi oscillations, while the longitudinal components imprint fast oscillating phases onto the target spin which have negligible effects on the target spin evolution in the weak-coupling regime ($A_z \leq A \ll \omega_0$). Thus Eq. (2) can be simplified as

$$H^{(\pm)}(t) = \pm \sum_{q=1}^{\infty} \frac{2(-1)^{q+1}\cos[(2q-1)\omega_T t]}{(2q-1)\pi}(A_\perp I_\perp) + \omega_0 I_z, \quad (3)$$

where $A_\perp = \sqrt{A_x^2 + A_y^2}$, and $I_\perp = \cos(\alpha)I_x + \sin(\alpha)I_y$ with $\alpha = \arctan(A_y/A_x)$. The phases of the transverse oscillating magnetic fields differ by $\pi$ for the two sensor states $|\pm\rangle$ [Fig. 2(a)]. Therefore, the target spin will be driven to evolve along opposite directions conditioned for the opposite sensor states [Figs. 2(b) and 2(c)], recording the which-way information of the sensor spin and hence causing the sensor spin decoherence.

Since the frequency differences between different harmonic components of the DD

filter function are much larger than the sensor-target coupling strength ($A \ll \omega_T$), so the effects of different harmonic components on the target spin evolution can be separately analysed. In particular for the first-order harmonic component of the DD filter function, the Hamiltonian in Eq. (3) can be simplified as $H^{(\pm)}(t) = \pm \frac{2A_\perp}{\pi}\cos(\omega_T t)I_\perp + \omega_0 I_z$. Now we can move to the rotating frame with respect to $\omega_T I_z$ and use the rotating wave approximation to get a time-independent Hamiltonian $H^{(\pm)} = \pm A_\perp I_\perp/\pi + (\omega_0 - \omega_T)I_z$, which is valid for the weak-coupling regime ($A_\perp \leq A \ll \omega_0$). The target spin performs Rabi oscillation with Rabi frequency $\omega_R = \sqrt{(A_\perp/\pi)^2 + (\omega_0 - \omega_T)^2}$. Then the sensor spin coherence is

$$L(t) \approx \frac{1}{2}\mathrm{Tr}\left[e^{iH^{(-)}t}e^{-iH^{(+)}t}\right] = 1 - \frac{2A_\perp^2}{\pi^2\omega_R^2}\sin^2\left(\frac{\omega_R t}{2}\right), \quad (4)$$

where we have assumed that initially the target spin is in the maximally mixed state. Since the sensor evolution time is $t = 2N\tau$, the sensor coherence in Eq. (4) is a function of both the pulse interval $\tau$ and the pulse number $N$,

$$L(N,\tau) \approx 1 - \frac{2A_\perp^2}{\pi^2\omega_R^2}\sin^2(N\tau\omega_R). \quad (5)$$

Note that $\omega_R$ depends on the pulse interval $\tau$ through $\omega_T$.

Conventional DD-based sensing schemes are based on identifying the "frequency fingerprints" of target spins. This is realized by sweeping the pulse interval with a constant pulse number to tune the effective AC magnetic fields into resonance with the target spin, i.e. $\omega_T = \omega_0$ for the first-order harmonic component so that the Rabi oscillation has the maximum amplitude (more generally, the resonance condition is $(2q-1)\omega_T = \omega_0$ with $q = 1, 2, \cdots$ denoting the order of sensor coherence dips). For CPMG control, $\omega_T = \pi/(2\tau)$, so the resonance condition is $2\tau = \pi/\omega_0$. Under this resonant DD condition, the target spin is resonantly driven to evolve along opposite directions for the opposite sensor states, causing maximum sensor spin decoherence [Fig. 2(c)]. Then by measuring the pulse interval corresponding to the sensor coherence dip, the Larmor frequency $\omega_0$ of the target spin can be determined. Such "frequency fingerprints" can be used to identify different spins and to determine the hyperfine interaction if it is strong enough to induce observable frequency shift of the target spin [10]. As shown in Fig. 2(d), for rather small CPMG pulse numbers, the coherence dip depth decreases from 1 to -1 as the pulse number increases, and the width of the sensor coherence dip is inversely proportional to the pulse number. As we increase the pulse number further, the sensor coherence at the dip position increases from -1 to 1 and shows strong side dips until the sensor spin coherence dip disappears completely. The appearance of strong side dips and smearing of the sensor coherence dips for large DD pulse numbers make it difficult to identify different target spins with slightly different Larmor frequencies [25].

The "wavefunction fingerprint" results from the sensitive dependence of the Rabi oscillation of the target spin on the hyperfine interaction. Even when the hyperfine interaction is too weak to induce any observable frequency shift of the target spin, the dependence of the Rabi oscillation of the target spin on the weak hyperfine interaction can still be revealed by observing the sensor coherence dip as a function of the DD pulse number with the pulse interval fixed for resonant DD ($2\tau = \pi/\omega_0$). Under the resonant DD condition ($\omega_0 = \omega_T$), the Rabi frequency $\omega_R = A_\perp/\pi$, so the sensor coherence dip as a function of the DD pulse number is obtained from Eq. (5) as

$$L_{\text{dip}}(N) \approx \cos\left(\frac{NA_\perp}{\omega_0}\right). \qquad (6)$$

This shows that the sensor coherence dip depth is a periodic function of the DD pulse number. Note that the sensor coherence dip as a function of DD pulse number has the same form as Eq. (6) if the initial state of the target spin is a pure state $|+1/2\rangle$ or $|-1/2\rangle$ (eigenstates of $I_z$). The physical picture is as follows. The sensor spin under DD control produces weak transverse oscillating magnetic fields, which have opposite phases for the opposite sensor states $|\pm\rangle$, causing the target spin to undergo opposite Rabi oscillations with the same Rabi frequency $A_\perp/\pi$. The sensor coherence dip is just the distance between the two bifurcated pathways of the target spin on the Bloch sphere, which oscillates periodically with the DD pulse number or the total evolution time [Fig.

2(e)].

The above discussion can be easily generalized to the case where a sensor spin is coupled to $M$ independent target spins of the same species ($I=1/2$), with the Hamiltonian

$$H(t) = f(t)S_z \otimes \sum_{k=1}^{M} \mathbf{A}_k \cdot \mathbf{I}_k + \omega_0 \sum_{k=1}^{M} I_k^z. \tag{7}$$

In quantum sensing, the coupling to the targets is weak ($A_k \ll \omega_0$). Keeping only the lowest-order harmonic component of $f(t)$, the Hamiltonian of the target spin conditioned on the sensor states is simplified as

$H^{(\pm)}(t) = \pm \dfrac{2}{\pi} \sum_{k=1}^{M} A_k^\perp \cos(\omega_T t) I_k^\perp + \omega_0 \sum_{k=1}^{M} I_k^z$, where $A_k^\perp = \sqrt{(A_k^x)^2 + (A_k^y)^2}$, and $I_k^\perp = \cos(\alpha_k) I_k^x + \sin(\alpha_k) I_k^y$ with $\alpha_k = \arctan(A_k^y/A_k^x)$. With the resonance condition satisfied, i.e. $\omega_T = \omega_0$ or $2\tau = \pi/\omega_0$, all the target spins are resonantly driven by the periodically modulated hyperfine coupling to the sensor spin, but with different Rabi frequencies due to different transverse coupling strengths $A_k^\perp$, so the sensor coherence dip as a function of the DD pulse number is

$$L_{\text{dip}}(N) \approx \prod_{k=1}^{M} \cos\left(\frac{NA_k^\perp}{\omega_0}\right), \tag{8}$$

the product of the contributions from different target spins. Although the coupling to each target spin to the sensor spin is weak ($A_k^\perp \ll \omega_0$), the transverse coupling strength of each target spin to the sensor spin can be quite different and results in different "wavefunction fingerprints" for different target spins, which causes characteristic sensor coherence dip oscillations as functions of the pulse number with the period $N_k = \pi\omega_0/A_k^\perp$ for the $k$ th target spin. If we can identify these pulse number periods from Eq. (8), we can accurately determine the transverse coupling strength $A_k^\perp$. The advantage of this new scheme is that it makes use of the relatively large "wavefunction gradients" of target spins induced by the sensor-target coupling gradients, instead of the slight "frequency gradients" of different target spins induced by the sensor spin in conventional sensing scheme [10-12].

## III. Resolving multiple spins of the same frequency

In this section, we demonstrate how to identify a few independent nuclear spins weakly coupled to a quantum sensor when the nuclear spins have the same "frequency fingerprint". We will study this issue using both the semiclassical noise model and the quantum model. We will show that in the semiclassical model the nuclear spins of the same frequency cannot be distinguished by the central spin decoherence, while in the quantum model it is feasible to resolve multiple nuclear spins of the same frequency by their different "wavefunction fingerprints". We assume that the quantum sensor is weakly coupled to $M$ distant target nuclear spins of the same species with the Hamiltonian given in Eq. (7).

### A. Semiclassical noise model

In the classical noise picture, the sensor accumulates random phases due to the magnetic noises produced by the environment and the target spins [28]. For an environment with a large number of particles, the magnetic noise can be approximated as a Gaussian stochastic process, which can be characterized by the two-point correlation function. For our model, the noise correlation function [29, 30] is defined as $C(t) = \langle \beta(t)\beta(0) \rangle = \langle e^{iH_0 t}\beta(0)e^{-iH_0 t}\beta(0) \rangle$, where $\beta = \sum_{k=1}^{M}(\mathbf{A}_k \cdot \mathbf{I}_k)$ is the noise operator, and $H_0 = \omega_0 \sum_{k=1}^{M} I_k^z$ is the free Hamiltonian for the target spins. We assume that the temperature is much higher than the target spin frequencies, i.e., the target spins are in the maximally mixed state. The correlation function is $C(t) = \frac{1}{4}\sum_{k=1}^{M}\left[\left(A_k^z\right)^2 + \left(A_k^\perp\right)^2 \cos(\omega_0 t)\right]$ and the noise power spectrum $S(\omega) = \frac{\pi}{4}\sum_{k=1}^{M}\left\{2\left(A_k^z\right)^2 \delta(\omega) + \left(A_k^\perp\right)^2 [\delta(\omega+\omega_0)+\delta(\omega-\omega_0)]\right\}$, containing a zero-frequency part (caused by the longitudinal components $\sum_{k=1}^{M} A_k^z I_k^z$ in the noise operator) and a finite-frequency part with the same frequency as the transitions of the target spins (caused by the transverse components $\sum_{k=1}^{M} A_k^\perp I_k^\perp$ in the noise operator).

Under DD control, the zero-frequency noise is eliminated and the sensor spin decoherence caused by the finite-frequency noise is [31]

$$L^c(t) = \exp\left[-\frac{1}{2}\int_{-\infty}^{\infty}\frac{d\omega}{2\pi}S(\omega)\frac{F^2(\omega,t)}{\omega^2}\right], \quad (9)$$

where $F(\omega,t) = \left|\sum_{p=0}^{N}(-1)^p(e^{i\omega t_{p+1}} - e^{i\omega t_p})\right|$ is the DD filter function ($t_0 = 0, t_{N+1} = t$). For the CPMG-$N$ control, the sensor coherence presents dips when the pulse interval matches the noise frequency, i.e., $2\tau_{\text{dip}} = \pi(2q-1)/\omega_0$ with $q = 1,2,\ldots$ denoting the dip order, as shown in Fig. 3(b). In the following we always consider the first-order coherence dip ($q=1$). At these dips (i.e., under the resonant DD condition), $F(\omega_0, t) = 2N$, so the sensor coherence dip depth is

$$L_{\text{dip}}^c(N) = \exp\left(-\sum_{k=1}^{M}(A_k^\perp)^2 \frac{N^2}{2\omega_0^2}\right), \quad (10)$$

with the total evolution time $t_{\text{dip}} = N\pi/\omega_0$ for resonant DD. In previous studies, the above formula has been used to extract the coupling strength between the sensor and target spins [10, 14]. However, the result in Eq. (10) is independent of the target spin number $M$ or the individual hyperfine coupling $A_k^\perp$ as long as the total noise strength $\sum_{k=1}^{M}(A_k^\perp)^2$ is kept constant. This means that in the Gaussian noise approximation one cannot distinguish whether the sensor coherence dips result from single or multiple target spins. The limitation of such a semiclassical model is that it cannot account for the manipulation of the target spins by the sensor spin and therefore misses the "wavefunction fingerprints" of the target spins, which are essential for resolving multiple target spins of the same species.

B. **Quantum model**

In the quantum model, the sensor spin decoherence is caused by the sensor-target entanglement when both the sensor spin and target spins are in pure states [32-34]. Under DD control, the sensor-target system can evolve with entanglement and disentanglement, leading to oscillation between decoherence and coherence recovery with increasing the DD pulse number [10, 20]. The entanglement is due to the target spin evolution conditioned on the sensor states $|\pm\rangle$, namely, $U_{(\pm)}^{N}(t) = e^{-i[H_0 \pm (-1)^N \beta/2](t-t_N)} \cdots e^{-i(H_0 \mp \beta/2)(t_2-t_1)} e^{-i(H_0 \pm \beta/2)t_1}$. In general the target spins are in the maximally mixed state and there is no authentic entanglement between the sensor spin and target spins. Yet the bifurcated quantum evolution of the target spins conditioned on the sensor state is still a useful picture to understand the sensor spin decoherence. Using the Magnus expansion [35], we obtain an approximate expression for the sensor coherence (see Appendix A)

$$L(t) = \frac{1}{2^M} \text{Tr}\left[\left(U_{(-)}^{N}\right)^{\dagger} U_{(+)}^{N}\right] \approx \prod_{k=1}^{M} \cos\left(\frac{A_k^{\perp}}{2\omega_0} F(\omega_0, t)\right), \tag{11}$$

for $M$ target spin-1/2's of the same frequency ($\omega_0$). In particular, when the period of the CPMG DD matches the noise frequency ($2\tau_{\text{dip}} = \pi/\omega_0$), the sensor coherence dip depth is

$$L_{\text{dip}}(N) \approx \prod_{k=1}^{M} \cos\left(\frac{NA_k^{\perp}}{\omega_0}\right), \tag{12}$$

which is the same as Eq. (8) in Sec. II. To check the validity of the above formula, we compare it to the exact numerical results in Fig. 3(c) and find excellent agreement in the weak coupling regime [$\sum_{k=1}^{M}\left(A_k^{\perp}\right)^2 \ll \omega_0$]. In the case of one target spin ($M=1$) the sensor coherence dip oscillates periodically between 1 and $-1$ with increasing the CPMG pulse number $N$. The oscillation period $N_k = 2\pi\omega_0/A_k^{\perp}$ is inversely proportional to the coupling strength $A_k^{\perp}$. For multiple target spins of the same noise frequency (but with different coupling strengths $A_k^{\perp}$'s to the sensor), the sensor coherence dip, being the product of $M$ different periodic oscillations with periods $N_k = 2\pi\omega_0/A_k^{\perp}$, oscillates non-periodically with $N$. For a large number of target spins ($M \gg 1$), the interference between oscillations of different periods $N_k = 2\pi\omega_0/A_k^{\perp}$ leads to a damped decay with increasing $N$, and the coherence dip depth is well described by the Gaussian noise approximation in Eq. (10) (see Appendix A).

The coupling strength $A_k^\perp$ of each target spin can be individually determined. The dip reaches zero whenever the pulse number is $N_k^{(0)} = \pi\omega_0/(2A_k^\perp)$ for a target spin. Therefore the number of zeros of the sensor coherence dip $N < N_{max}$ determines the number of target spins coupled to the sensor with coupling strength above a threshold $A_k^\perp > \pi\omega_0/(2N_{max})$; the positions of the zeros $N_k^{(0)} = \pi\omega_0/(2A_k^\perp)$ determine quantitatively the individual coupling strengths. The zeros of the sensor coherence dip $L_{dip}(N)$ are indeed clearly seen in Fig. 3(d). Thus by identifying the zeros of the sensor coherence dip as a function of DD pulse number, we can resolve, in a discrete manner, multiple target spins even when they have the same noise frequency.

## IV. Characterization of spin correlations

### A. Identifying different types of nuclear spin transitions

Here we show that the correlations in a single spin cluster weakly coupled to a sensor can be characterized, in a discrete manner, by observing the sensor coherence dip as a function of the DD pulse number. Let us first consider how to distinguish two specific types of target spin clusters, namely, type-II and type-V [Fig. 4(a)], referred to as uncorrelated and correlated, respectively. Type-II transitions are produced by two independent target spin-1/2's $\mathbf{I}_A$ and $\mathbf{I}_B$; type-V transitions are produced by a correlated cluster (such as a spin-1 $\mathbf{J}$ formed by two interacting spin-1/2's). The Hamiltonians for the sensor coupled to these two types of target spins are

$$H_{II} = \lambda_{II} S_z \left( I_A^x + I_B^x \right) + \omega_A I_A^z + \omega_B I_B^z, \tag{13}$$

$$H_V = \lambda_V S_z J_x + \frac{\omega_A + \omega_B}{2} J_z^2 + \frac{\omega_A - \omega_B}{2} J_z, \tag{14}$$

where the transitions frequencies $\omega_{A/B}$ are set the same for the type-II and type-V

transitions and the couplings to the sensor $\lambda_V = \sqrt{3}\lambda_{II}/2$. The parameters are chosen such that the two types of target spins produce identical noise power spectrum to the sensor. The conventional noise spectroscopy cannot distinguish the correlated and the uncorrelated transitions.

However, the target spin clusters of different types of correlations would perform qualitatively different quantum evolutions under the modulated hyperfine interaction and thus induce different behaviors of sensor-target entanglement and disentanglement under DD control. In particular, the sensor coherence dip as a function of the CPMG pulse number $N$ presents different oscillation features for different types of correlations [Fig. 4(b)]. Again, by the Magnus expansion [35] (see Appendix B), we get

$$L^{II}(t) \approx \cos\left(\frac{\lambda_{II}}{2\omega_A} F(\omega_A, t)\right)\cos\left(\frac{\lambda_{II}}{2\omega_B} F(\omega_B, t)\right), \tag{15}$$

$$L^{V}(t) \approx \frac{1}{3}\left[1 + 2\cos\left(\lambda_V \sqrt{\frac{F^2(\omega_A, t)}{2\omega_A^2} + \frac{F^2(\omega_B, t)}{2\omega_B^2}}\right)\right]. \tag{16}$$

Without loss of generality, we assume $\omega_B \neq \omega_A$. Then in the case of DD resonant with transition $A$, i.e., $2\tau_{\text{dip}} = \pi/\omega_A$, the DD filter functions $F(\omega_A, t_{\text{dip}}) = 2N$ and $F(\omega_B, t_{\text{dip}}) \ll F(\omega_A, t_{\text{dip}})$ for large $N$, so the sensor coherence dips for the two types of target spins are

$$L_{\text{dip}}^{II}(N) \approx \cos(\lambda_{II} N/\omega_A), \tag{17}$$

$$L_{\text{dip}}^{V}(N) \approx \frac{1}{3}\left[1 + 2\cos\left(\sqrt{2}\lambda_V N/\omega_A\right)\right]. \tag{18}$$

The sensor coherence dips oscillate with different amplitudes for the two different types of target spin correlations: For type-II correlations (independent target spins), the minimum sensor coherence dip is $-1$, but for type-V transitions (correlated target spins), the minimum sensor coherence dip is $-1/3$. The coherence dip minima take discretely different values for different types of correlations in the target spins.

As a more general case, we show that the DD quantum sensing can distinguish the transitions of independent spin-1/2's and those in a higher spin [ladder-type in Fig. 4(a)], even if the two types of target systems have the same noise power spectrum. We assume the coupling between a sensor and a target spin with spin $J$ has the Hamiltonian $H = \lambda S_z J_x + \sum_{m=-J}^{J} \varepsilon_m |m\rangle\langle m|$ (where the basis states are assumed to satisfy $J_z |m\rangle = m|m\rangle$). The noise spectrum caused by the target spin is

$$S(\omega) = \frac{2\pi}{2J+1} \sum_{m=-J}^{J-1} \lambda_m^2 \left[\delta(\omega+\omega_m) + \delta(\omega-\omega_m)\right]$$ where the transition $|m\rangle \leftrightarrow |m+1\rangle$

causes the noise with frequency $\omega_m = |\varepsilon_{m+1} - \varepsilon_m|$ with the amplitude $\lambda_m = \lambda\sqrt{(J-m)(J+m+1)}/2$. Then when the DD is resonant with a transition ($2\tau_{\text{dip}} = \pi/\omega_m$), the sensor coherence dip as a function of the DD pulse number $N$ is (see Appendix C)

$$L_{\text{dip}}^{(m)}(N) = \frac{2J-1}{2J+1} + \frac{2}{2J+1} \cos\left(\frac{2\lambda_m N}{\omega_m}\right). \tag{19}$$

Here we have assumed the general case that different transitions have different frequencies such that if $F(\omega_m, t_{\text{dip}}) = 2N$ then $F(\omega_n, t_{\text{dip}}) \approx 0$ for $n \neq m$. Thus a spin-$J$, or a ladder-type system with $2J+1$ levels, is characterized by a discrete depth of the sensor coherence dip, that is

$$\min\left(L_{\text{dip}}^{(m)}\right) = \frac{2J-3}{2J+1}, \tag{20}$$

which occurs at the characteristic DD pulse number $N_m = \pi\omega_m/(2\lambda_m)$. Such discrete features of the sensor coherence dip caused by correlated transitions of target spins are clearly seen in Fig. 4(c).

The above results for type-II, type-V and ladder-type correlations can be easily extended to the most general case of a bonded nuclear spin cluster weakly coupled to a spin-1/2 sensor. If this nuclear spin cluster has $d$ eigenstates denoted by $\{|m\rangle\}$, then when the DD is resonant with a specific transition $|m\rangle \leftrightarrow |n\rangle$ the sensor coherence dip has a discrete minimum of sensor coherence dip, $\min\left(L_{\text{dip}}^{(m,n)}\right) = (d-4)/d$ (see Appendix C). The discrete minima can be explained by a physical picture similar to that in Sec. II. If we only consider the subspace of the cluster $\{|m\rangle, |n\rangle\}$, which is resonant with the DD frequency, the situation is analogous to that for a spin-1/2 target spin and the sensor coherence dip will oscillate between 1 and -1. But the probability for the cluster in the maximally mixed state to be in such a subspace is $2/d$. The cluster has also the probability $(d-2)/d$ to be in the other states, which are not affected by the DD control and for which the sensor coherence is constantly 1. So the sensor coherence dip minimum is $-2/d + (d-2)/d = (d-4)/d$. Thus using these discrete values of the sensor coherence dip, we can characterize, in a discrete manner, the correlation size of target nuclear spins.

**B. Distinguishing independent nuclear spins from a nuclear spin cluster**

Let us specifically consider a nuclear spin cluster containing two nuclear spins with different gyromagnetic ratios with the Hamiltonian

$$H_C = \lambda S_z \left(I_A^x + I_B^x\right) + \omega_A I_A^z + \omega_B I_B^z + 2\mu I_A^z I_B^z. \tag{21}$$

Note that Eq. (21) differs from Eq. (13) for type-II transition in the extra coupling term $2\mu I_A^z I_B^z$. The energy levels of the nuclear spin cluster is shown in Fig. 5(a). If these two nuclear spins are nearly independent, the degenerate transitions $|+1/2\rangle_A |+1/2\rangle_B \leftrightarrow |-1/2\rangle_A |+1/2\rangle_B$ and $|+1/2\rangle_A |-1/2\rangle_B \leftrightarrow |-1/2\rangle_A |-1/2\rangle_B$ with transition energy $\omega_A$ correspond to the transition $|+1/2\rangle_A \leftrightarrow |-1/2\rangle_A$ in type-II [Fig. 4(a)]. These two transitions will both contribute to the sensor coherence dip if the DD pulse interval matches the transition energy [$\tau = \pi/(2\omega_A)$], so the sensor coherence dip as a function of the DD pulse number is the same as that for a single target spin 1/2, i.e. the coherence dip oscillates between 1 and -1. However, the coupling between these two nuclear spins breaks the degeneracy of the two transitions by inducing a transition energy splitting $2\mu$. If the energy splitting is much larger than the sensor-target coupling strength, the DD set resonant with the transition $|+1/2\rangle_A |+1/2\rangle_B \leftrightarrow |-1/2\rangle_A |+1/2\rangle_B$ (with frequency $\omega_A + \mu$) would not induce the other transition ($|+1/2\rangle_A |+1/2\rangle_B \leftrightarrow |-1/2\rangle_A |+1/2\rangle_B$ (with frequency $\omega_A - \mu$). This is the case for a correlated spin cluster with Hilbert space dimension $d = 4$ and the sensor coherence dip oscillates between 1 and 0. Numerical results show that the two transitions can be separated as long as $\mu \geq \lambda$ [Fig. 5(b)].

## V. MRI of spin-labeled single molecules

We now demonstrate that the coherence dip features of a quantum sensor under DD control can be employed for angstrom-resolution MRI of single molecules. We take a shallow NV center in diamond near the surface as the sensor and consider the MRI of single trimethylphosphite (TMP) molecules with $^{31}$P nuclear spins as labels and the conformation analysis of a single Villa (2F4K) protein molecule [36] labeled by $^{15}$N nuclear spins in the PHE amino acids. We assume that the diamond is $^{13}$C-depleted [10] (with $^{13}$C abundance 0.01%) so that the effect of the background $^{13}$C nuclear spins on the sensing is negligible [25].

The Hamiltonian for an NV electron spin weakly coupled to multiple nuclear spins is

$$H = S_z \sum_{k=1}^{M} \mathbf{A}_k \cdot \mathbf{I}_k - \gamma_n \sum_{k=1}^{M} \mathbf{B} \cdot \mathbf{I}_k + H_{\text{dip}}, \quad (22)$$

where the NV spin-1 $S_z$ has eigenstates $\{|0\rangle, |\pm 1\rangle\}$, $\mathbf{A}_k$ is the hyperfine field for the $k$ th target spin, $\mathbf{B}$ is the external magnetic field, and $\gamma_n$ is the gyromagnetic ratio of the target spins, and $H_{\text{dip}} = \sum_{i<j} D_{ij} \left[ \mathbf{I}_i \cdot \mathbf{I}_j - 3(\mathbf{I}_i \cdot \mathbf{r}_{ij})(\mathbf{r}_{ij} \cdot \mathbf{I}_j)/r_{ij}^2 \right]$ is the dipolar interaction between the nuclear spins, with $D_{ij} = \mu_0 \gamma_n^2 / (4\pi r_{ij}^3)$ and $\mu_0$ denoting the vacuum permeability. The hyperfine interaction between the NV sensor and the $k$ th target spin is $A_k \sim \mu_0 \gamma_e \gamma_n / (4\pi R_k^3)$ with $R_k$ being the distance from the NV sensor to the $k$ th target spin. For two $^{31}$P nuclear spins $i$, $j$ with $R_i, R_j \sim 3$ nm and $r_{ij} \sim 0.3$ nm, the sensor-target coupling strength is about two times larger than the nuclear spin dipolar interaction, and for two $^{15}$N nuclear spins under the same condition, the sensor-target coupling strength is more than six times larger than the nuclear spin dipolar interaction. So the dipolar interaction between the target spins has negligible effects on the sensor decoherence and the target spins can be assumed to be independent of each other, as numerically demonstrated [25].

Here we choose the transition between the basis states $\{|+1\rangle, |-1\rangle\}$ of the NV center as the sensor, to ensure that the target nuclear spins have the same Larmor frequency. Similar to Eq. (12), the NV electron spin coherence dip as a function of CPMG pulse number is $L_{\text{dip}}(N) \approx \prod_k^M \cos(2A_k^\perp N/\omega_0)$ (see Appendix D), where $A_k^\perp$ is the component of the hyperfine field perpendicular to the magnetic field, and $\omega_0 = |\gamma_n \mathbf{B}|$ is the Larmor frequency of target nuclear spins.

With a single TMP molecule on the diamond surface [Fig. 6(a)], the oscillations of the NV coherence dip with the CPMG pulse number have different periods for different magnetic field directions [Fig. 6(b)], which determines $A_k^\perp$, the component of the hyperfine field orthogonal to the magnetic field. By measuring the periods of the sensor coherence dip oscillations for three different magnetic field directions, the hyperfine interaction can be fully reconstructed (see Appendix E) and therefore the position of the target $^{31}$P nuclear spin in the TMP can be precisely determined [Fig. 6(c)]. With three TMP molecules on the diamond surface [Fig. 6(d)], the NV coherence dip shows non-periodic oscillations as the CPMG pulse number is increased [Fig. 6(e)], with the coherence dip zeros corresponding to the couplings $A_k^\perp$ of different $^{31}$P nuclear spins to the NV center. By choosing three different magnetic field directions [Fig. 6(f)], we can individually determine the couplings and hence the positions of the $^{31}$P nuclear spins (Table I). In the simulations, the three TMP molecules are about 2.2~2.5 nm away from the NV sensor and are placed in a line with the minimum distance between each other as 0.3 nm, so the spatial resolution can reach 3 Å.

The DD-based MRI can be applied to analyze the conformations of bio-molecules. In a 2F4K protein molecule that has 35 residues, for example, we can label the four

PHE amino acids with $^{15}$N and perform the MRI of the $^{15}$N nuclear spins and hence analyze the conformation of the protein molecule. In Figs. 6(g)-(i), we show that the positions of the four $^{15}$N nuclear spin labels can be precisely determined by using the oscillation features of the NV center spin coherence dip under resonant DD control (Table I). In the simulations, the four $^{15}$N nuclear spin labels are located about 2.5~3 nm away from the NV sensor and the distance between different spin labels are about 0.6~1.5 nm.

## VI. Discussion

Besides the DD-based quantum sensing, there are other sensing methods, such as the cross-polarization method [37-39] and the correlation spectroscopy method [40, 41]. The cross-polarization method tunes the sensor spin and target spins into the Harmann-Hahn double resonance and then extracts the sensor-target coupling strength from the time-domain polarization transfer between the sensor spin and target spins. This method requires relatively strong coupling between the sensor spin and a single target spin and cannot distinguish multiple target spins with nanoscale resolution. We notice that a recent adaption of the cross-polarization method [39] proposes to use an ancillary nuclear spin near the sensor as a quantum memory to significantly increase the spatial resolution. The two-dimensional (2D) spectroscopy method proposed in Ref. [39] relies on polarizing the target nuclear spins, which is analogous to the conventional 2D NMR [3]. The correlation spectroscopy method transfers the NV electron spin phase that contains the information about the target spins to the NV spin population and therefore greatly increases the spectral resolution by relying on the long $T_1$ time of the NV center. This method has been used to detect the Larmor precession of statistically polarized $^{13}$C [40] and $^1$H nuclear spin ensembles [41]. However, the correlation spectroscopy schemes are not designed for resolving nuclear spins with the same Larmor frequency or distinguishing uncorrelated and correlations nuclear spin transitions.

The wavefunction fingerprint sensing scheme in this paper can not only resolve multiple nuclear spins of the same species, but also distinguish the uncorrelated and correlations nuclear spin transitions. As compared with other methods, it has several advantages: (i) It works in the weak sensor-target coupling regime, which is especially relevant in quantum sensing; (ii) the target spins can be at the infinite-temperature state (i.e., no polarization is needed); (iii) it does not need specific control over the target spins or an ancillary quantum memory. The discrete wavefunction fingerprints provide a further advantage: Small errors in measurement can be well tolerated since the correlations are distinguished by the discrete minimum values of the sensor coherence dip (e.g., the dip minima $-1/3 \pm 0.05$ and $0 \pm 0.05$ are separated by more than 6 standard deviations).

Now we estimate the measurement time for the MRI scheme based on the NV center in this paper. The coherence of the double transition ($|+1\rangle \leftrightarrow |-1\rangle$) of the NV center can be generated, controlled and read out in a way similar to the coherence of the single transitions ($|0\rangle \leftrightarrow |\pm 1\rangle$) by using composite microwave pulses [42, 43]. The signal-to-noise ratio (SNR) for $K$ measurements of the NV spin coherence, with the spin projection noise and photon shot noise considered, is $\sigma = 1/(F\sqrt{K})$ [19], where $F = \left[1 + 2(\alpha_0 + \alpha_1)/(\alpha_0 - \alpha_1)^2\right]^{-1/2}$ is the readout fidelity with $\alpha_0$, $\alpha_1$ being the mean number of detected photons per shot from the $|0\rangle$ and $|\pm 1\rangle$ states of the NV center, respectively. To reach a desired SNR $\sigma$, the measurement cycle need to be repeated for $K = 1/(F\sigma)^2$ times. The total measurement time for observing the sensor coherence dip zero caused by the $k$ th target spin is $T_k = K(t_k^D + t_{IR})$, where $t_k^D = \pi^2/(4A_k^\perp)$ is the time for the evolution time under DD control and $t_{IR} \sim 1$ μs is the time for the initialization and readout of the NV center. For a TMP molecule placed about 3 nm away from the NV center, $A_k^\perp/(2\pi) \sim 1$ kHz, $t_k^D = \pi^2/(4A_k^\perp) \approx 0.4$ ms. For typical fluorescence collection efficiencies, the NV readout fidelity is $F \approx 0.03$. To reach the standard error $\sigma = 0.01$, the measurement cycle need to repeated for $K \approx 1.1 \times 10^7$ times and the total measurement time is $T_k \approx 4.4 \times 10^3$ s. Recently the NV readout fidelity has been improved to about $F \approx 0.3$ by storing the NV electron spin state in an ancillary $^{15}$N nuclear spin [44]. In this case, only $1.1 \times 10^5$ measurements are needed and the total measurement time is reduced to about 44 s.

The MRI scheme requires that the NV electron spin relaxation time $T_1$ and the coherence time $T_2$ of a near-surface NV center be in the millisecond time scale. For a slowing-fluctuating environment (such as the dilute $^{13}$C nuclear spin in isotope-purified diamond), the DD control largely suppresses the transverse spin decoherence and the sensor coherence can be well preserved before the sensor spin relaxation happens [25]. In this sense, our sensing scheme is limited by the $T_1$ time. But for a fast-fluctuating environment (such as the electron spin bath which may exist in diamond surface), the DD control is ineffective in suppressing the bath noise and therefore the sensing scheme is limited by the $T_2$ time. For shallow NV centers, the fluctuating surface electron spins are the main noise source [45-50]. For an NV center 2~4 nm below the surface, $T_1$ at room temperature is limited to 430~960 $\mu$s, and $T_2$ is within 5~10 $\mu$s for Hahn echo and can be prolonged to more than 50 $\mu$s under multi-pulse DD control [45]. We expect that in the future it may be possible to further extend both $T_1$ and $T_2$ by material engineering [47, 48] or lowering the temperature [49, 51]. We note that recently the coherence time $T_2$ of a shallow NV center has been prolonged to about 100 $\mu$s through a surface-treatment technique [44]. In the short term, the scheme can be applied to NV centers inside bulk with the electron spin coherence reaching a few milliseconds as limited by the spin relaxation time $T_1 \approx 7.7$ ms at room temperature [52]. It may also be possible to use the near-surface NV centers to detect electron spin labels in single molecules [53], since compared with the nuclear spins, the electron spins have much stronger interaction with NV sensor spins, therefore requiring much shorter detection time.

## VII. Conclusion

DD-based quantum sensing has enabled atomic-scale NMR by identifying the "frequency fingerprints" of target spins. In this paper, we provide a conceptual advance from the "*frequency fingerprints*" to the "*wavefunction fingerprints*" in DD-based quantum sensing. We provide a solution to distinguishing spins or transitions that have the same frequencies but have different wavefunction evolutions, using the fact that the quantum evolutions of target spins under resonant DD are much more sensitive to the weak hyperfine interaction than the frequencies of target spins are. We also find a surprising effect - the amplitudes of the sensor coherence oscillations under resonant DD have discrete values, which makes identification of different spin numbers or different types of correlations robust against small measurement errors. The scheme in this work offers an approach to breaking the resolution limit set by the "frequency gradients" in conventional MRI and hence to reaching the angstrom-scale resolution in quantum sensing.

## Acknowledgements

This work was supported by Hong Kong Research Grants Council - Collaborative Research Fund CUHK4/CRF/12G and the Chinese University of Hong Kong Vice Chancellor's One-off Discretionary Fund.

## Appendix A: Derivation of Eq. (12) in the main text

The Hamiltonian representing a spin-1/2 sensor interacting with $M$ spin-1/2 target spins is

$$H = S_z \sum_{k=1}^{M} (A_k^x I_k^x + A_k^y I_k^y + A_k^z I_k^z) + \omega_0 \sum_{k=1}^{M} I_k^z. \tag{A1}$$

It can be written in the eigenbasis of the sensor $\{|\pm\rangle\}$ as

$$H = |+\rangle\langle+| \otimes H^{(+)} + |-\rangle\langle-| \otimes H^{(-)}, \tag{A2}$$

with the Hamiltonian of the target spins conditioned on the sensor state

$$H^{(\pm)} = \pm \frac{1}{2}\beta + H_0, \tag{A3}$$

where $\beta = \sum_{k=1}^{M}(A_k^x I_k^x + A_k^y I_k^y + A_k^z I_k^z)$ is the noise operator for the sensor, and $H_0 = \omega_0 \sum_{k=1}^{M} I_k^z$ is the free Hamiltonian for the target spins. In the interaction picture set by $H_0$, the time-dependent noise operator is

$$\beta(t) = e^{iH_0 t} \beta e^{-iH_0 t}$$
$$= \sum_{k=1}^{M} \left\{ \frac{A_k^{\perp}}{2} \left[ I_k^+ e^{i(\omega_0 t - \alpha_k)} + I_k^- e^{-i(\omega_0 t - \alpha_k)} \right] + A_k^z I_k^z \right\}, \tag{A4}$$

where $A_k^{\perp} = \sqrt{(A_k^x)^2 + (A_k^y)^2}$ and $\alpha_k = \arctan(A_k^y / A_k^x)$. The time evolution operator of the target spins dependent on the sensor state is

$$U_{(\pm)}^N(t) = e^{-iH_0 t} \hat{T} e^{\mp \frac{i}{2} \int_0^t f(t') \beta(t') dt'}, \tag{A5}$$

where $f(t) = (-1)^p$ for $[t_p, t_{p+1}]$ is the DD modulation function ($t_0 = 0, t_{N+1} = t$) and $\hat{T}$ is the time-ordering operator.

Now we use the Magnus expansion [35] to get a simple formula for $U_N^{(\pm)}(t)$, which is valid for weak sensor-target coupling. According to the Magnus expansion, a general time-dependent evolution operator can expanded as

$$U(t) = \hat{T} e^{-i \int_0^t H(t') dt'} = \exp\left( \sum_{l=1}^{\infty} \Omega_l(t) \right), \tag{A6}$$

with the first-order and second-order Magnus terms

$$\Omega_1(t) = -i \int_0^t H(t') dt', \tag{A7}$$

$$\Omega_2(t) = -\frac{1}{2} \int_0^t dt_1 \int_0^{t_1} dt_2 [H(t_1), H(t_2)]. \tag{A8}$$

For our specific model, the first-order Magnus term is

$$\Omega_1(t) = -\frac{i}{4} \sum_{k=1}^{M} A_k^{\perp} \int_0^t f(t') \left[ I_k^+ e^{i(\omega_0 t' - \alpha_k)} + I_k^- e^{-i(\omega_0 t' - \alpha_k)} \right] dt'$$
$$= -i \sum_{k=1}^{M} \frac{A_k^{\perp}}{2\omega_0} F(\omega_0, t) I_k^{(\xi - \alpha_k)}. \tag{A9}$$

where

$$F(\omega, t) = \omega \left| \int_0^t f(t') e^{i\omega t'} dt' \right|, \tag{A10}$$

$$e^{i\xi} = \left( \omega \int_0^t f(t') e^{i\omega t'} dt' \right) \Big/ F(\omega, t), \tag{A11}$$

$$I_k^\xi = \cos\xi I_k^x - \sin\xi I_k^y. \tag{A12}$$

Note that the time average of $I_k^z$ is zero due to the balanced filter function $f(t)$. The second-order Magnus term is

$$\Omega_2(t) = -i\sum_{k=1}^{M} \frac{\left(A_k^\perp\right)^2}{2\omega_0^2} \chi_2(\omega,t) I_k^z, \tag{A13}$$

where

$$\chi_2(\omega,t) = \omega^2 \int_0^t dt_1 \int_0^{t_1} dt_2 f(t_1) f(t_2) \sin[\omega(t_1 - t_2)]. \tag{A14}$$

In the weak-coupling regime $A_k^\perp \ll \omega_0$, it is obvious that $\|\Omega_2(t)\| \ll \|\Omega_1(t)\|$. Moreover, at the resonance points of $\omega$, i.e. $t = \pi(2q-1)N/\omega$ ($q = 1, 2, \cdots$), it is easy to numerically verify that $\Omega_2(t) \approx 0$ [see Fig. 4 in Ref. 35 for the numerical comparison of $F(\omega,t)/(\omega t)$ and $|\chi_2(\omega,t)|/(\omega t)^2$]. So we can use the first-order Magnus expansion to approximate $U_{(\pm)}^N(t)$ as

$$\begin{aligned} U_{(\pm)}^N(t) &\approx \exp(-iH_0 t)\exp[\pm\Omega_1(t)] \\ &= \exp(-iH_0 t)\exp\left[\mp i\sum_{k=1}^{M} \frac{A_k^\perp}{2\omega_0} F(\omega_0,t) I_k^{(\xi-\alpha_k)}\right], \end{aligned} \tag{A15}$$

and the product of $\left(U_{(-)}^N(t)\right)^\dagger$ and $U_{(+)}^N(t)$ as

$$\begin{aligned} \left(U_{(-)}^N(t)\right)^\dagger U_{(+)}^N(t) &\approx \exp\left[-i\sum_{k=1}^{M} \frac{A_k^\perp}{\omega_0} F(\omega_0,t) I_k^{(\xi-\alpha_k)}\right] \\ &= \prod_{k=1}^{M} \left[\cos\varphi_k - 2i\sin\varphi_k I_k^{(\xi-\alpha_k)}\right], \end{aligned} \tag{A16}$$

where $\varphi_k = \frac{A_k^\perp}{2\omega_0} F(\omega_0,t)$. So the qubit coherence is

$$L(t) = \frac{1}{2^M}\text{Tr}\left[\left(U_{(-)}^N(t)\right)^\dagger U_{(+)}^N(t)\right] \approx \prod_{k=1}^{M} \cos\left[\frac{A_k^\perp}{2\omega_0} F(\omega_0,t)\right]. \tag{A17}$$

For the CPMG-$N$ control, the filter function is [31]

$$F(\omega,t)=\begin{cases} 4\sin^2\left(\dfrac{\omega t}{4N}\right)\left|\cos\left(\dfrac{\omega t}{2}\right)\cos^{-1}\left(\dfrac{\omega t}{2N}\right)\right|, \text{odd } N, \\ 4\sin^2\left(\dfrac{\omega t}{4N}\right)\left|\sin\left(\dfrac{\omega t}{2}\right)\cos^{-1}\left(\dfrac{\omega t}{2N}\right)\right|, \text{even } N. \end{cases} \quad (A18)$$

At the resonance points $t_{\text{dip}} = \pi(2q-1)N/\omega_0$ ($q=1,2,\cdots$), $F(\omega_0, t_{\text{dip}}) = 2N$, so the sensor coherence dip can be further simplified as

$$L_{\text{dip}}(N) \approx \prod_{k=1}^{M} \cos\left(\frac{A_k^\perp N}{\omega_0}\right), \quad (A19)$$

which is Eq. (12) in the main text. For small CPMG pulse numbers, the sensor coherence dip from the quantum model agrees with the Gaussian noise approximation,

$$\begin{aligned} L_{\text{dip}}(N) &\approx 1 - \sum_{k=1}^{M} \left(A_k^\perp\right)^2 \frac{N^2}{2\omega_0^2} \\ &\approx \exp\left(-\sum_{k=1}^{M}\left(A_k^\perp\right)^2 \frac{N^2}{2\omega_0^2}\right). \end{aligned} \quad (A20)$$

## Appendix B: Derivation of Eq. (16) in the main text

The Hamiltonian for a spin-1/2 sensor coupled to a spin-1 target spin is

$$H_V = \lambda_V S_z J_x + \frac{\omega_A + \omega_B}{2} J_z^2 + \frac{\omega_A - \omega_B}{2} J_z. \quad (B1)$$

The noise operator is $\beta = \lambda_V J_k^x$ for the sensor. The free Hamiltonian of the target spin is $H_0 = \dfrac{\omega_A + \omega_B}{2} J_z^2 + \dfrac{\omega_A - \omega_B}{2} J_z$. In the interaction picture set by $H_0$, the time-dependent noise operator is

$$\beta(t) = \frac{\lambda_V}{\sqrt{2}} \begin{bmatrix} 0 & e^{i\omega_A t} & 0 \\ e^{-i\omega_A t} & 0 & e^{-i\omega_B t} \\ 0 & e^{i\omega_B t} & 0 \end{bmatrix}. \quad (B2)$$

So the first-order Magnus term is

$$\Omega_1(t) = -\frac{i\lambda_V}{2\sqrt{2}} \begin{bmatrix} 0 & \frac{F(\omega_A,t)e^{i\xi_A}}{\omega_A} & 0 \\ \frac{F(\omega_A,t)e^{-i\xi_A}}{\omega_A} & 0 & \frac{F(\omega_B,t)e^{-i\xi_B}}{\omega_B} \\ 0 & \frac{F(\omega_B,t)e^{i\xi_B}}{\omega_B} & 0 \end{bmatrix}, \quad (B3)$$

where $e^{i\xi_{A/B}} = \left(\omega_{A/B} \int_0^t f(t')e^{i\omega_{A/B}t'} dt'\right) \Big/ F(\omega_{A/B}, t)$, and we have used $F(\omega,t) = F(-\omega,t)$.

The sensor coherence is

$$\begin{aligned} L^V(t) &= \frac{1}{3}\text{Tr}\left[\exp(2\Omega_1(t))\right] \\ &= \frac{1}{3}\left[1 + 2\cos\left(\lambda_V \sqrt{\frac{F^2(\omega_A,t)}{2\omega_A^2} + \frac{F^2(\omega_B,t)}{2\omega_B^2}}\right)\right], \end{aligned} \quad (B4)$$

which is Eq. (16) in the main text.

## Appendix C: Derivation of Eq. (19) in the main text

We consider a general Hamiltonian of a spin-1/2 sensor coupled to a target spin cluster,

$$H = S_z \beta + H_0, \quad (C1)$$

with

$$H_0 = E_n \sum_{n=1}^d |n\rangle\langle n|, \quad (C2)$$

$$\beta = \frac{1}{2}\sum_{m,n}\left(\beta_{mn}|m\rangle\langle n| + \text{H.c.}\right), \quad (C3)$$

where $d$ is the number of eigenstates of the target spin cluster. Then the time-dependent noise operator is

$$\beta(t)=\frac{1}{2}\sum_{m,n}\left(\beta_{mn}e^{i\omega_{mn}t}|m\rangle\langle n|+\text{H.c.}\right). \tag{C4}$$

So the first-order Magnus term is

$$\Omega_1(t)=-\frac{i}{2}\sum_{m>n}\left(\beta_{mn}\frac{F(\omega_{mn},t)e^{i\xi_{mn}}}{\omega_{mn}}|m\rangle\langle n|+\text{H.c.}\right), \tag{C5}$$

where $e^{i\xi_{mn}}=\left(\omega_{mn}\int_0^t f(t')e^{i\omega_{mn}t'}dt'\right)/F(\omega_{mn},t)$. Here the diagonal terms with $m=n$ in $\beta(t)$ are averaged out by the DD control, so there is no diagonal terms in $\Omega_1(t)$. Without loss of generality, we assume the transitions have different frequencies. The sensor coherence dip due to the transition of frequency $\omega_{mn}$ is

$$L_{\text{dip}}^{(mn)}(N)=\frac{1}{d}\text{Tr}\left[\exp(2\Omega_1(t))\right]=\frac{1}{d}\left[d-2+2\cos\left(2\left|\frac{\beta_{mn}}{\omega_{mn}}\right|N\right)\right]. \tag{C6}$$

Here we have assumed that different transition frequencies have no joint contributions to the same sensor coherence dip, that is, if $F(\omega_{mn},t_{\text{dip}})=2N$, then $F(\omega_{pq},t_{\text{dip}})\approx 0$ for $p,q\neq m,n$, which can be satisfied if different transitions have different frequencies and the DD pulse number $N$ is large. Then there are only two non-zero eigenenergies of $2\Omega_1(t_{\text{dip}})$: $\pm 2i\left|\frac{\beta_{mn}}{\omega_{mn}}\right|N$. Therefore we have the above formula for $L_{\text{dip}}^{(mn)}(N)$.

In particular, for a target spin-$J$ with sensor-target coupling $\lambda$,

$$H=\lambda S_z J_x+\sum_{m=-J}^{J}\varepsilon_m|m\rangle\langle m|. \tag{C7}$$

For the target spin transition $|m\rangle\leftrightarrow|m+1\rangle$, we have $\lambda_m=\lambda\sqrt{(J-m)(J+m+1)}/2$, so the sensor coherence dip is

$$L_{\text{dip}}^{(m)}(N)=\frac{2J-1}{2J+1}+\frac{2}{2J+1}\cos\left(\frac{2\lambda_m N}{\omega_m}\right), \tag{C8}$$

where $\omega_m=|\varepsilon_{m+1}-\varepsilon_m|$. This formula is just Eq. (19) in the main text.

## Appendix D: MRI via an NV center in diamond

The Hamiltonian for an NV electron spin (S=1) coupled to M nuclear spins (I=1/2) is

$$H = S_z \sum_{k=1}^{M} \mathbf{A}_k \cdot \mathbf{I}_k - \gamma_n \sum_{k=1}^{M} \mathbf{B} \cdot \mathbf{I}_k. \quad (D1)$$

For the sake of simplicity, we assume that the magnetic field is along the z direction (it is easy to generalize the following derivation to the magnetic field along any direction). Then the Hamiltonian becomes

$$H = S_z \sum_{k=1}^{M} (A_k^x I_k^x + A_k^y I_k^y + A_k^z I_k^z) + \omega_0 \sum_{k=1}^{M} I_k^z, \quad (D2)$$

where $\omega_0 = -\gamma_n B$ (here we assume $\gamma_n < 0$; for $\gamma_n > 0$, define $\omega_0 = \gamma_n B$ and change Eq. (D2) correspondingly). This Hamiltonian can be written in the eigenbasis of the NV electron spin $\{|0\rangle, |\pm 1\rangle\}$ as

$$H = |+1\rangle\langle +1| \otimes H^{(+1)} + |-1\rangle\langle -1| \otimes H^{(-1)} + |0\rangle\langle 0| \otimes H^{(0)}, \quad (D3)$$

with the Hamiltonian of the target spins conditioned on the sensor state

$$H^{(\eta)} = \eta \, \beta + H_0, \quad (D4)$$

where $\eta = -1, 0,$ or $+1$, $\beta = \sum_{k=1}^{M}(A_k^x I_k^x + A_k^y I_k^y + A_k^z I_k^z)$ is the noise operator for the sensor, and $H_0 = \omega_0 \sum_{k=1}^{M} I_k^z$ is the free Hamiltonian of the target spins. In this paper, we choose the basis $\{|+1\rangle, |-1\rangle\}$ to sense the target spins. In the interaction picture set by $H_0$, the time-dependent noise operator is

$$\beta(t) = e^{iH_0 t} \beta e^{-iH_0 t}$$
$$= \sum_{k=1}^{M} \left\{ \frac{A_k^{\perp}}{2} \left[ I_k^+ e^{i(\omega_0 t - \alpha_k)} + I_k^- e^{-i(\omega_0 t - \alpha_k)} \right] + A_k^z I_k^z \right\}, \quad (D5)$$

where $A_k^{\perp} = \sqrt{(A_k^x)^2 + (A_k^y)^2}$ and $\alpha_k = \arctan(A_k^y / A_k^x)$. Then the first-order Magnus term is

$$\Omega_1(t) = -i \sum_{k=1}^{M} \frac{A_k^\perp}{\omega_0} F(\omega_0, t) I_k^{(\xi-\alpha_k)}, \tag{D6}$$

where $F(\omega_0, t)$, $\xi$, and $I_k^\xi$ are correspondingly defined in Eq. (A10) and (A11), and (A12). It is straightforward to obtain the following equation in the main text,

$$L_{\text{dip}}(N) \approx \prod_{k=1}^{M} \cos\left(\frac{2 A_k^\perp N}{\omega_0}\right). \tag{D7}$$

## Appendix E: Scheme for reconstruction of target spin locations

In Sec. V of the main text, we employ an NV center as the sensor to perform MRI of multiple nuclear spins with the same Larmor frequency. Here we give details for the reconstruction of target spins locations. We denote the position of the $k$th target spin relative to the NV sensor as $\mathbf{R}_k = (R_k^x, R_k^y, R_k^z) = R_k \mathbf{n}_k$ with $\mathbf{n}_k = (r_k^x, r_k^y, r_k^z)$ and $r_k^\alpha = R_k^\alpha / R_k$ ($\alpha = x, y, z$). The Hamiltonian for an NV center weakly coupled to multiple nuclear spins is

$$H = S_z \sum_k^M \mathbf{A}_k \cdot \mathbf{I}_k - \gamma_n \sum_k^M \mathbf{B} \cdot \mathbf{I}_k + H_{\text{dip}}, \tag{E1}$$

where $S_z$ is the NV electron spin-1 operator (we choose the states $\{|+1\rangle, |-1\rangle\}$ as the sensor), $\mathbf{I}_k$ is the target nuclear spin operator, $\mathbf{A}_k = A_k \mathbf{a}_k$ is the hyperfine field for the $k$th nuclear spin with the hyperfine field strength

$$A_k = \frac{\mu_0 \gamma_e \gamma_n}{R_k^3} \sqrt{1 + 3(n_k^z)^2}, \tag{E2}$$

the unit vector of the hyperfine field

$$\mathbf{a}_k = \frac{\left(-3 n_k^x n_k^z, -3 n_k^y n_k^z, 1 - 3(n_k^z)^2\right)}{\sqrt{1 + 3(n_k^z)^2}}, \tag{E3}$$

and $\mathbf{B} = B\mathbf{m}$ is the external magnetic field with $\mathbf{m}$ being the unit vector and $\gamma_n$ the gyromagnetic ratio of the target spins.

The sensor spin coherence dip as a function of the CPMG pulse number $N$ is $L_{\text{dip}} \approx \prod_k^M \left(2 A_k^\perp N / \omega_0\right)$, where $\omega_0 = |\gamma_n| B$ is the target spin Larmor frequency and $A_k^\perp = A_k \sqrt{1 - (\mathbf{m} \cdot \mathbf{a}_k)^2}$ is the component of the hyperfine field perpendicular to the magnetic field. If the pulse number period $N_k = \pi \omega_0 / A_k^\perp$ is measured for the $k$th target spin, the perpendicular hyperfine field strength $A_k^\perp$ is determined. But the obtained $A_k^\perp$ for a single magnetic field direction cannot fully determine the position $\mathbf{R}_k$ of the $k$th target spin. We can perform the measurements for three different magnetic field directions $\{\mathbf{m}_1, \mathbf{m}_2, \mathbf{m}_3\}$ with the same magnetic field strength, and obtained three pulse number periods $\{N_k^1, N_k^2, N_k^3\}$. Then we have the following set of equations,

$$\begin{cases} A_k \sqrt{1 - (\mathbf{m}_1 \cdot \mathbf{a}_k)^2} = \pi \omega_0 / N_k^1, \\ A_k \sqrt{1 - (\mathbf{m}_2 \cdot \mathbf{a}_k)^2} = \pi \omega_0 / N_k^2, \\ A_k \sqrt{1 - (\mathbf{m}_3 \cdot \mathbf{a}_k)^2} = \pi \omega_0 / N_k^3. \end{cases} \quad (E4)$$

By solving these equations, we first obtain the unit vector of the hyperfine field vector $\mathbf{a}_k$. The distance $R_k$ and the unit direction vector $\mathbf{n}_k$ of the $k$th target spin are further determined by Eq. (E2) and (E3).

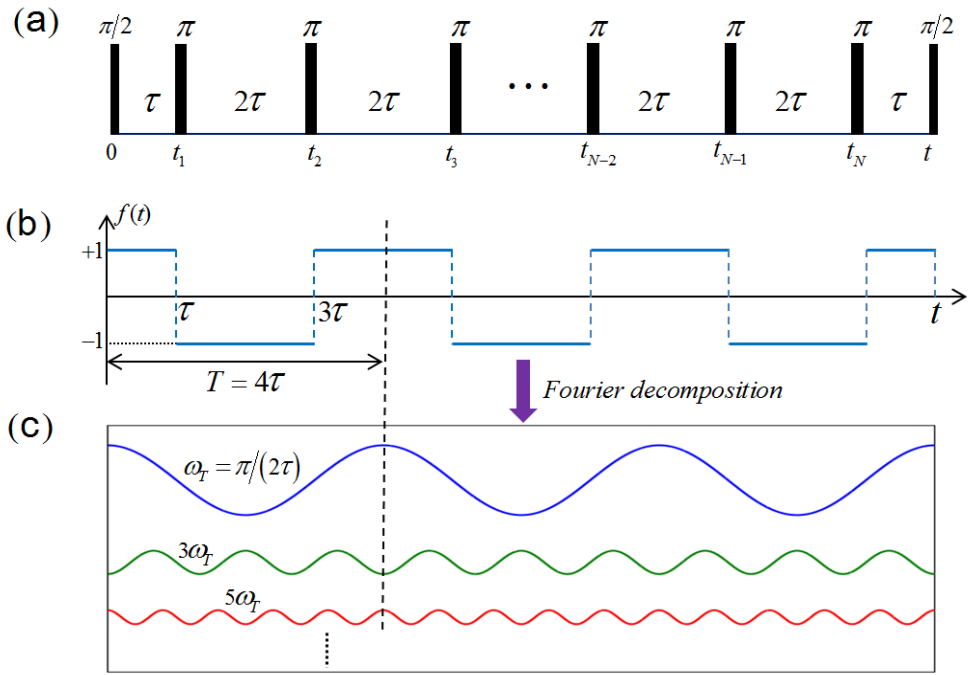

**Fig. 1.** (a) Schematic of an $N$-pulse CPMG control sequence with pulse interval $2\tau$ and pule frequency $\omega_T = \pi/2\tau$. (b) The modulation function of a 6-pulse CPMG sequence (CPMG-6). (c) Fourier decomposition of the modulation function in (b) into harmonics with discrete frequencies $(2q-1)\omega_T$ for $q = 1, 2, \cdots$.

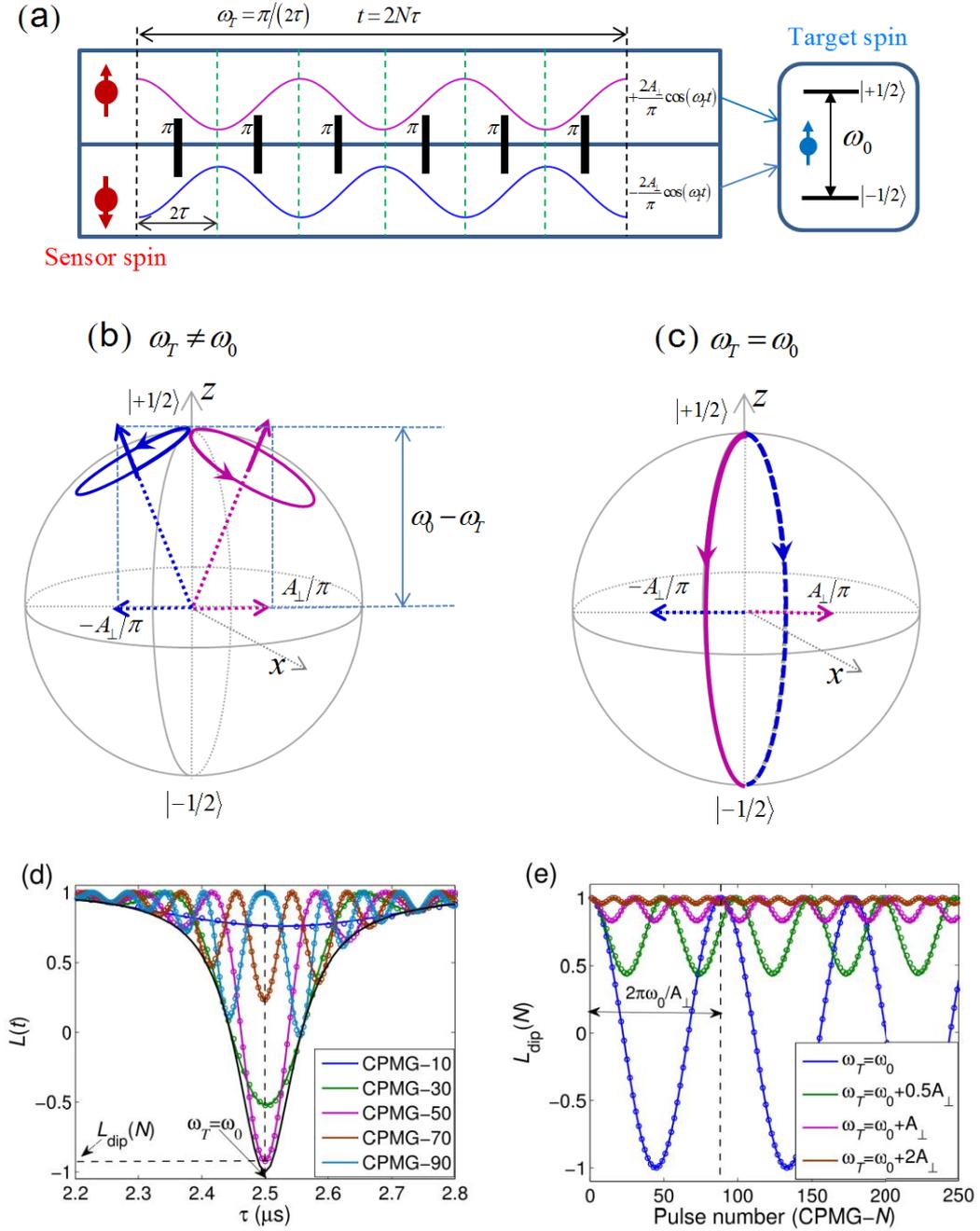

**Fig. 2.** (a) Schematic of the oscillating transverse magnetic fields produced by the sensor spin under CPMG control. The phase of the oscillating field depends on the initial state of the sensor spin. Here we only consider the lowest-order harmonic component of the DD filter function with frequency $\omega_T$. (b,c) Precession of the target spin conditioned on the sensor spin state in the rotating reference frame with respect to $\omega_T I_z$, with the CPMG pulse frequency off-resonant ($\omega_T \neq \omega_0$) and resonant ($\omega_T = \omega_0$) with the target spin Larmor frequency. Here we have assumed that the initial state of the target spin is $|+1/2\rangle$. (d) Sensor spin coherence as a function of the CPMG pulse interval for various CPMG pulse numbers. The black line is the sensor coherence envelope $L(\tau) \approx 1 - 2A_\perp^2/(\pi^2 \omega_R^2)$. (e) Sensor coherence dip as a function of the CPMG pulse number for pulse sequences resonant and off-resonant with the target spin Larmor frequency. In (d) and (e), the symbols are exact results while the solid lines are the analytical results from Eq. (5). The parameters are such that the target spin Larmor frequency $\omega_0 = 0.1$ MHz and the coupling strengths $A_x = A_y = A_z = 5$ kHz ($A_\perp = 7.07$ kHz).

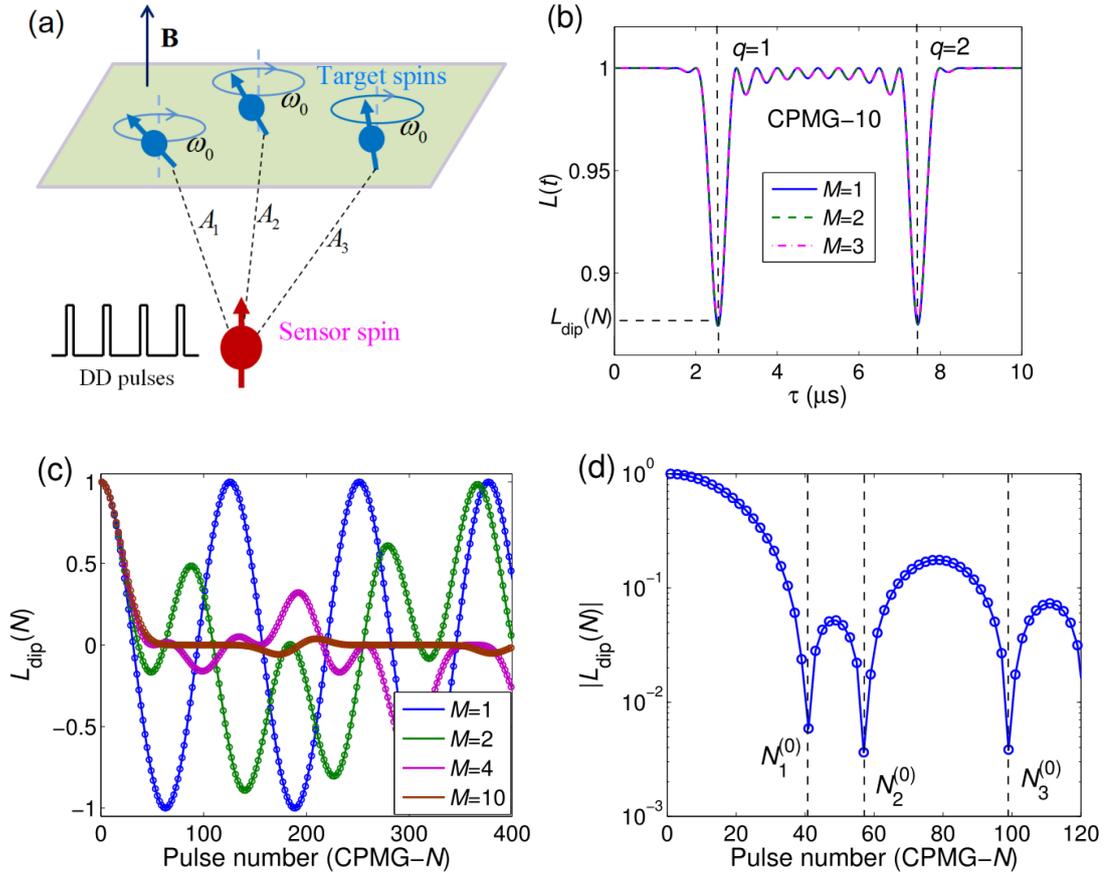

**Fig. 3.** Resolution of multiple nuclear spins of the same species weakly coupled to a quantum sensor. (a) Schematic of detecting multiple distant target spins of the same species by a sensor spin. The target spins precess with the same Larmor frequency and are not distinguished in their noise spectra. (b) Sensor coherence caused by various numbers of target nuclear spins as a function of time under the 10-pulse CPMG DD control (CPMG-10). The sharp dips correspond to the DD timing resonant with the noise frequency of the nuclear spins (resonant DD), with $q$ denoting the coherence dip order. The parameters are such that the nuclear spin precession frequency $\omega_0 = 0.1$ MHz and the coupling strengths $\sum_{k=1}^{M}(A_k^{\perp})^2 = 5$ kHz (with the coupling coefficients $A_k^{\perp}$ randomly chosen within the constraint) and $A_k^z = A_k^{\perp}/\sqrt{2}$. (c) Sensor coherence dip as a function of the CPMG DD pulse number $N$ for various numbers of target spins ($M=1, 2, 4, 10$). Symbols are the exact results from the quantum model while solid lines are the analytical results from Eq. (12). (d) Logarithm plot of the absolute value of the sensor coherence dip caused by 3 target spins ($M=3$) as a function of the CPMG pulse number.

The coherence dip zeros determine the number of target spins $M$ and the individual coupling strengths $A_k^\perp = \pi \omega_0 / (2 N_k^{(0)})$, where $N_k^{(0)}$ is the pulse number for which the sensor coherence dip is zero.

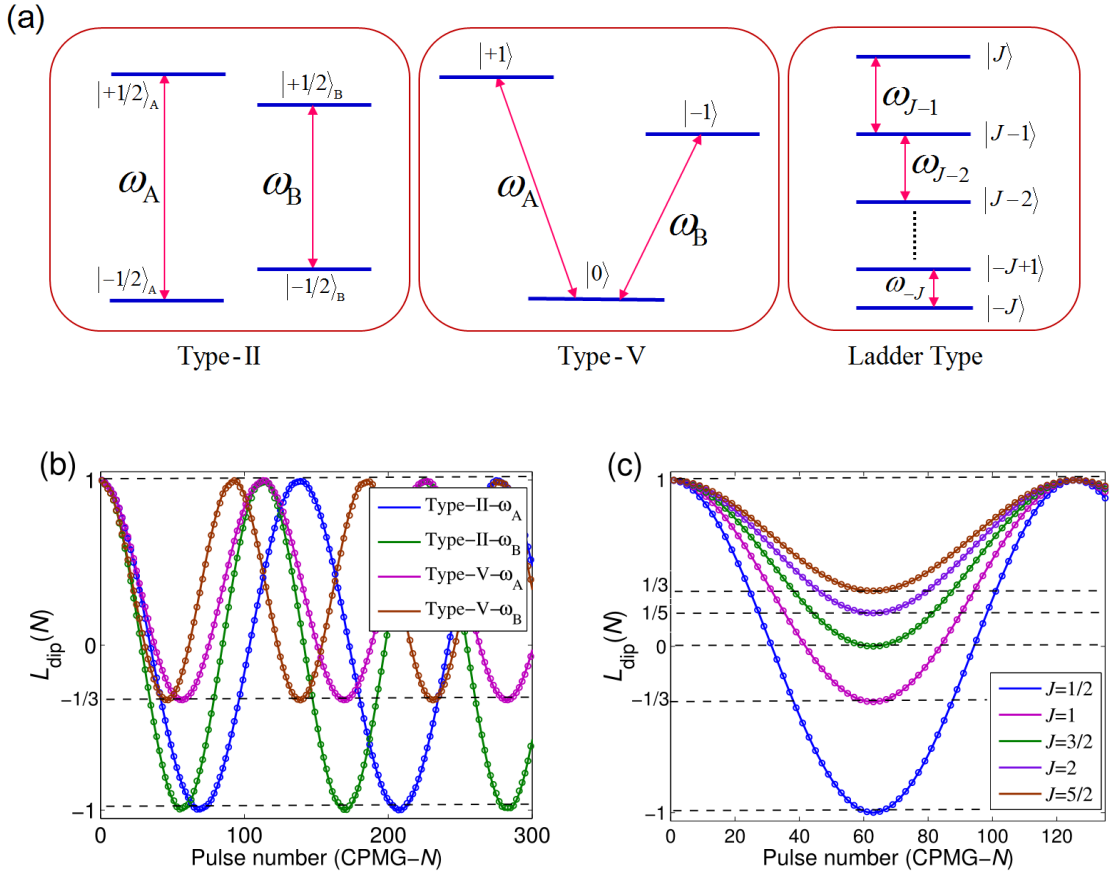

**Fig. 4.** Discrete features of sensor coherence dips due to different types of correlations of nuclear spins. (a) Schematic of different types of correlations of nuclear spins. Type-II transitions represent two independent target spin-1/2's, type-V transitions represent a single target spin-1, and ladder-type transitions represent a single target spin-$J$ or a strongly-bonded spin cluster with dimension $2J+1$. (b) Sensor coherence dip as a function of the CPMG pulse number for type-II and type-V transitions. The two type of transitions are set to have the same noise spectrum. Symbols are the exact results from the quantum model while solid lines are the results from the analytical formula in Eq. (17) and (18). The parameters such that are $\omega_A = 0.11$ MHz, $\omega_B = 0.09$ MHz, $\lambda_{II} = 5$ kHz, and $\lambda_V = \sqrt{3}\lambda_{II}/2$. (c) Sensor coherence dip as a function of the CPMG pulse number, for ladder-type transitions with various numbers of levels ($2J+1$). The DD timing is fixed to be resonant with one noise frequency. Symbols are the exact results from quantum model while solid lines are the results from the analytical formula

in Eq. (19). The transition frequency resonant with DD is $\omega_{J-1} = 0.1$ MHz and the coupling to the sensor is $\lambda_{J-1} = 2.5$ kHz.

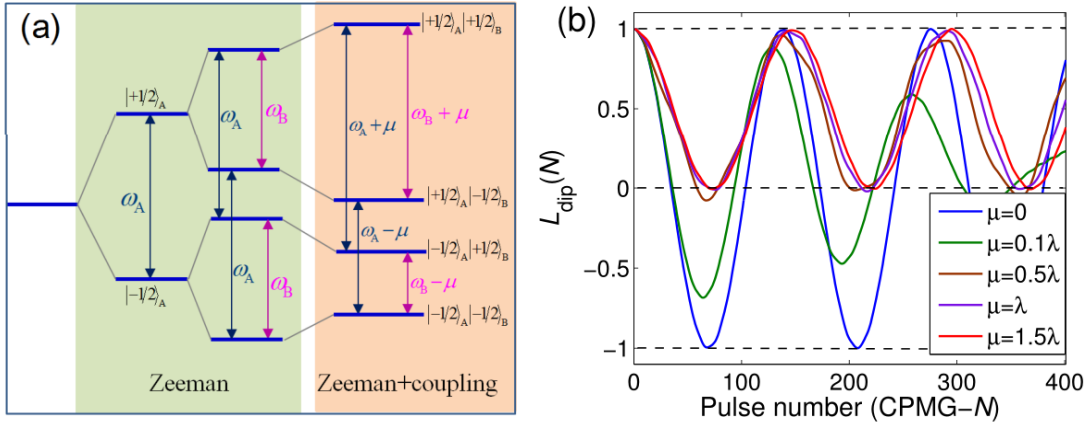

**Fig. 5.** Distinguishing independent nuclear spins from a nuclear spin cluster. (a) Energy levels for a pair of nuclear spins with different Larmor frequencies without and with coupling to each other. Without coupling, the energy level diagram is equivalent to that of the type-II system in Fig. 4(a). (b) Sensor coherence dip caused by the nuclear spin pair as a function of the CPMG pulse number for various coupling strengths. Without coupling ($\mu = 0$), we choose the transition $|+1/2\rangle_A \leftrightarrow |-1/2\rangle_A$ with transition frequency $\omega_A$; with coupling ($\mu \neq 0$), we choose the transition $|+1/2\rangle_A |+1/2\rangle_B \leftrightarrow |-1/2\rangle_A |+1/2\rangle_B$ with transition frequency $\omega_A + \mu$. The Hamiltonian of the nuclear spin pair is shown in Eq. (21) with the parameters $\omega_A = 0.11$ MHz, $\omega_B = 0.09$ MHz, and $\lambda = 5$ kHz.

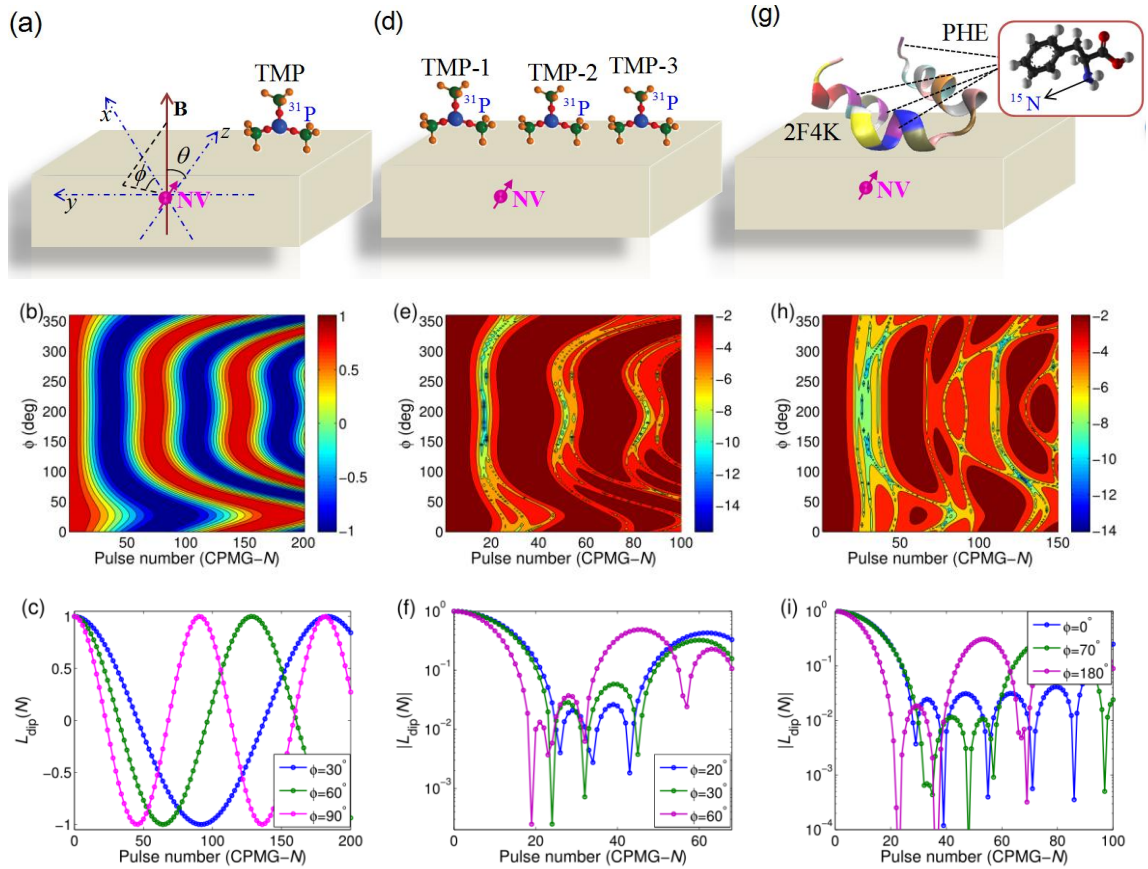

**Fig. 6.** Angstrom-resolution MRI of nuclear spin labels in single molecules. (a) Schematic of detecting the $^{31}$P nuclear spin in a TMP molecule by a shallow NV center (2 nm from the surface) in a $^{13}$C-depleted diamond. The NV axis along the [111] crystal direction is set as the $z$ axis, and the magnetic field has the polar angle $\theta$ and azimuth angle $\phi$ (relative to the crystal axis $[\bar{1}\bar{1}2]$). (b) Contour plot of the sensor coherence dip as a function of the CPMG pulse number and the azimuth angle $\phi$ for a 50 Gauss magnetic field with $\theta=40°$. (c) Sensor coherence dip as a function of the CPMG pulse number for three different magnetic field directions (for three values of $\phi$ with fixed $\theta=40°$). (d) Schematic of MRI of three TMP molecules near the diamond surface using $^{31}$P labelling. (e) Similar to (b) but here the contour represents the logarithm of the absolute value of the coherence dip. (f) Similar to (b) but here for three TMP molecules. (g) Schematic of conformation analysis of a 2F4K protein molecule via MRI of four $^{15}$N nuclear spin labels in the PHE amino acids (with residue sequence numbers 6, 10, 17 and 35). (h), (i) similar to (e), (f) respectively, but for the four $^{15}$N nuclear spins in the 2F4K protein.

**Table I** | Positions of nuclear spin labels in three TMP molecules on a diamond surface and in a 2F4K protein molecule on a diamond surface. The data are obtained by the DD-based MRI. The real positions are shown in comparison. The uncertainties in the positions obtained by the DD-based MRI originate from the SNR in determining the sensor coherence dip zeros, which is set as 0.01 in the simulations.

| Nuclear spin labels | Real positions (x, y, z) (Å) | Positions by DD-based MRI (x, y, z) (Å) |
|---|---|---|
| TMP-1-$^{31}$P | (13.88, 7.07, 18.48) | (12.88$\pm$1.06, 7.25$\pm$0.34, 18.22$\pm$0.18) |
| TMP-2-$^{31}$P | (15.11, 4.95, 16.74) | (15.18$\pm$3.53, 3.67$\pm$3.56, 16.91$\pm$0.84) |
| TMP-3-$^{31}$P | (16.33, 2.83, 15.01) | (14.26$\pm$2.40, 3.83$\pm$2.73, 15.36$\pm$0.44) |
| 2F4K-R06-$^{15}$N | (12.58, 13.40, 17.32) | (13.45$\pm$2.62, 13.19$\pm$3.40, 17.35$\pm$1.00) |
| 2F4K-R10-$^{15}$N | (16.75, 12.37, 21.78) | (16.85$\pm$1.02, 12.40$\pm$1.11, 21.70$\pm$0.62) |
| 2F4K-R17-$^{15}$N | (20.90, 6.01, 16.25) | (20.87$\pm$1.01, 4.09$\pm$4.00, 16.26$\pm$0.65) |
| 2F4K–R35-$^{15}$N | (9.09, 4.67, 24.54) | (11.66$\pm$1.82, 6.32$\pm$0.38, 23.70$\pm$0.58) |